\documentclass[pra, aps,reprint,superscriptaddress]{revtex4-2}         

\usepackage{graphicx}
\usepackage{dcolumn}
\usepackage{bm}
\usepackage{amsmath}

\newcommand{\NF}{\mathrm{N_F}} 
\newcommand{\NS}{\mathrm{N_S}}

\begin{document}


\title{Direct visualization of local electric fields in a layer of a ferroelectric nematic liquid} 

\author{Anej Sterle}
\email{anej.sterle@ijs.si}
\affiliation{Jožef Stefan Institute, Ljubljana, Slovenia}
\affiliation{Faculty of Mathematics and Physics, University of Ljubljana, Ljubljana, Slovenia}

\author{Natan Osterman}
\affiliation{Jožef Stefan Institute, Ljubljana, Slovenia}
\affiliation{Faculty of Mathematics and Physics, University of Ljubljana, Ljubljana, Slovenia}

\author{Calum J. Gibb}
\affiliation{School of Chemistry, University of Leeds, Leeds, UK}

\author{Jordan Hobbs}
\affiliation{School of Physics and Astronomy, University of Leeds, Leeds, UK}

\author{Richard J. Mandle}
\affiliation{School of Chemistry, University of Leeds, Leeds, UK}
\affiliation{School of Physics and Astronomy, University of Leeds, Leeds, UK}

\author{Nerea Sebastián}
\affiliation{Jožef Stefan Institute, Ljubljana, Slovenia}

\author{Alenka Mertelj}
\affiliation{Jožef Stefan Institute, Ljubljana, Slovenia}

\date{\today}

\begin{abstract}
Ferroelectric nematic liquids exhibit complex ferroelectric domains shaped by competing elastic and electrostatic interactions. Using fluorescence and polarizing optical microscopy, we investigate domain formation and evolution under different anchoring conditions. Charged fluorescent ions map the electrostatic potential, revealing that electric fields are localized near domain walls and material surfaces. Upon cooling, domain walls transform from Ising- to Néel-type configurations, lowering the electrostatic potential, and reversibly recover upon heating.
\end{abstract}

\maketitle

\textit{Introduction}---The discovery and development of ferroelectric nematic liquids \cite{mandle_rational_2017,nishikawa_fluid_2017} have revealed new and interesting phenomena related to ferroelectricity. Ferroelectric nematic liquids are soft ferroic materials with spontaneous polarization values comparable to those of solid ferroelectrics. Consequently, the electrostatic contribution to the free energy plays a dominant role, leading to the formation of characteristic domain structures and unique field responses \cite{barboza_explosive_2022,sterle_light-induced_2023,caimi_fluid_2023,cmok_running_2023,mathe_electric_2023,mathe_electrically_2024}. While electrostatic energy minimization drives domain formation, domain structuring is strongly influenced by surface conditions. The spontaneous polarization \textbf{P} is parallel to the nematic director field \textbf{n}, and anchoring strategies such as rubbing and photoalignment, which are commonly used for conventional nematic liquid crystals, can be employed \cite{caimi_surface_2021,sebastian_polarization_2023}. However, since the presence of the spontaneous polarization breaks the inversion symmetry of the nematic director, the boundary conditions can also be polar. Experimentally, it has been shown that the rubbing direction that induces a certain pretilt of the preferred director orientation at the surface causes a polar boundary condition \cite{caimi_surface_2021,chen_polar_2021}, while the photoalignment provides a nonpolar boundary condition \cite{sebastian_polarization_2023}. 

The domains are separated by the domain walls, where the \textbf{P} direction inverts or changes by a certain angle over a short distance.  One of the simplest ways the inversion of \textbf{P} can occur is through an Ising wall, in which \textbf{P} first decreases to zero and then increases to -\textbf{P}. However, deep in the ferroelectric nematic phase ($\mathrm{N_F}$), the domain walls are more complex; typically, the inversion walls and walls dividing $\pi$-twisted domains consist of one or more topological defect lines in the \textbf{n} field \cite{yi_chiral_2024,lavrentovich_twist_2025,sebastian_polarization_2023}. 

Despite the anticipated importance of electrostatic effects in governing domain morphology and stability, direct experimental measurements of the associated local electric fields have remained challenging. Here, we show how fluorescent ions can be used to evaluate local electric fields in ferroelectric nematic liquids. We exploited this to demonstrate how electrostatics governs the domain and domain walls evolution. In the equilibrated structures, the domain structure is such that the local fields are mostly compensated everywhere but at the sample surface and in the narrow regions around domain walls. Furthermore, we show that electrostatic forces drive topological transformations of the domain walls and the annihilation of the domains. 

\textit{Experimental}---To probe local fields, a small amount of modified Rhodamine B (mRhB)(\cite{erkoreka_flexoelectricity-driven_2026}, End matter) is dissolved in the liquid crystal mixture F7 with the phase sequence N 100°C $\NS$ 91°C $\NF$(\cite{gibb_spontaneous_2024}, End Matter). By fluorescence microscopy (FM), the distribution of mRhB fluorescent cations is probed. In the presence of the spatially dependent electrostatic potential $\phi(\mathbf{r})$, the concentration of cations will be larger in the regions, where $\phi(\mathbf{r})$ is smaller. Away from the sample surfaces and phase boundaries, a Boltzmann distribution of ions can be assumed $c^+=c_0 e^{-\phi_n}$, where $\phi_n=e\phi/(k_BT)$, \textit{e} the cation charge and $k_BT$ the thermal energy. In such a case, the fluorescence intensity at a given position in the image is $I(x,y)\propto \int_0^d c^+ (x,y)\,dz$, and the average intensity of the image is  $I_{avg}\propto c_0 d$, where \textit{d}  is the sample thickness. The average normalized 2D potential can be approximated  as 
\begin{equation} 
\begin{split}
 \qquad \langle\phi_n \rangle=-\log(I/I_{avg})
\end{split}
\label{eq1}
\end{equation}
and, if $|\phi_n| \ll 1$, $\langle \phi_n \rangle \approx 1-\frac{I}{I_{avg}}$. At the phase boundaries, differences in solvation energy (ion partitioning) can lead to a higher concentration of ions in one phase \cite{everts_ionically_2021}.

\begin{figure*}
\includegraphics[width=1\textwidth]{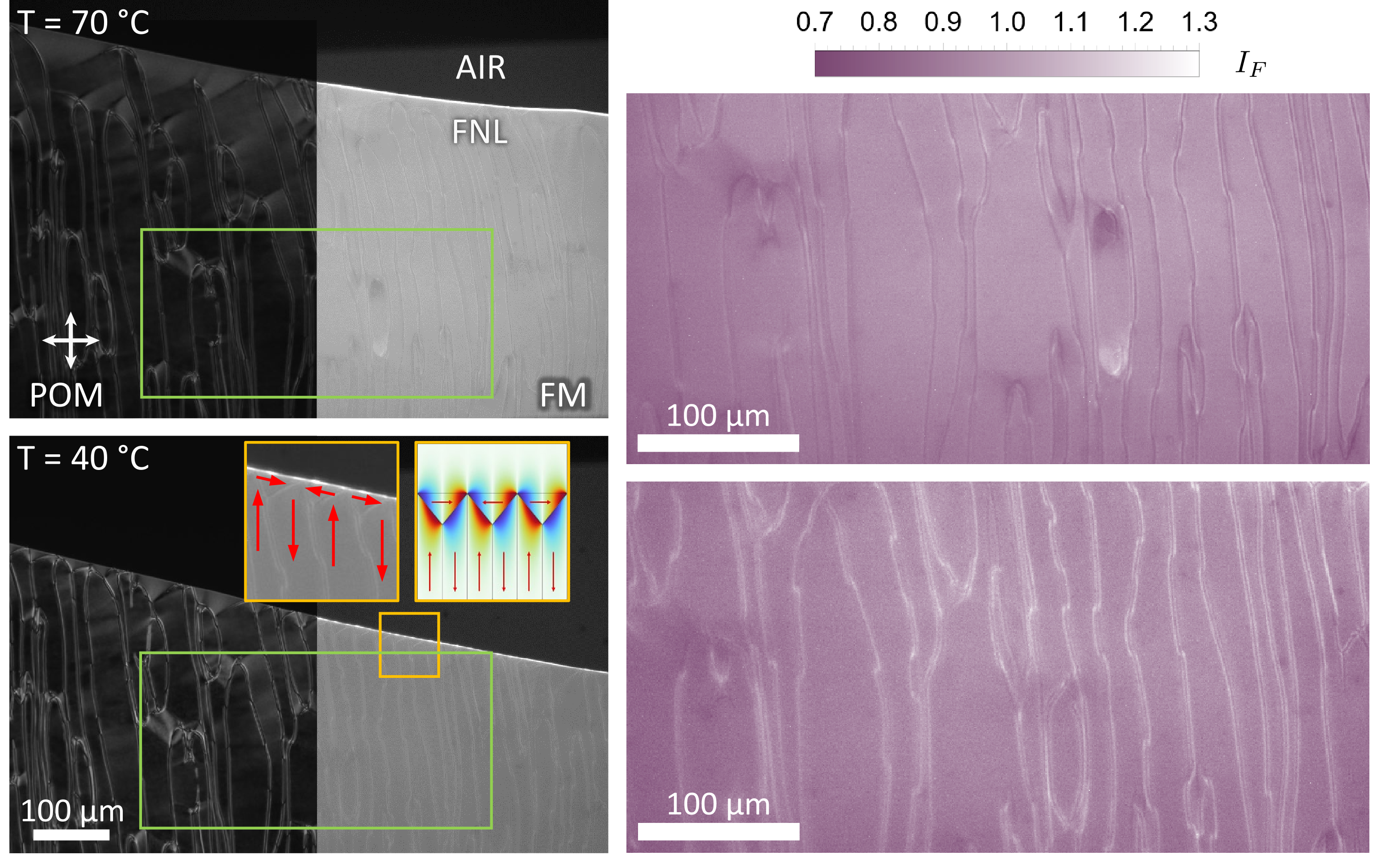}
\caption{\label{fig:fig1} Polarizing optical (POM) and fluorescence microscopy (FM) images of ferroelectric domains in a 5 $\mu$m thick partially filled LC cell with F7 + 0.008 wt\% mRhB mixture without a surface treatment layer at $70^{\circ}$C and $40^{\circ}$C. A close-up of ferroelectric domains at the LC–air interface is shown in orange, with red arrows indicating the polarization direction. The corresponding electric potential is shown schematically, with red representing positive and blue representing negative potential. Normalized FM images at different temperatures are shown on the right, with $I_F=I/I_{avg}$.}
\end{figure*}

\begin{figure*}
\includegraphics[width=1\textwidth]{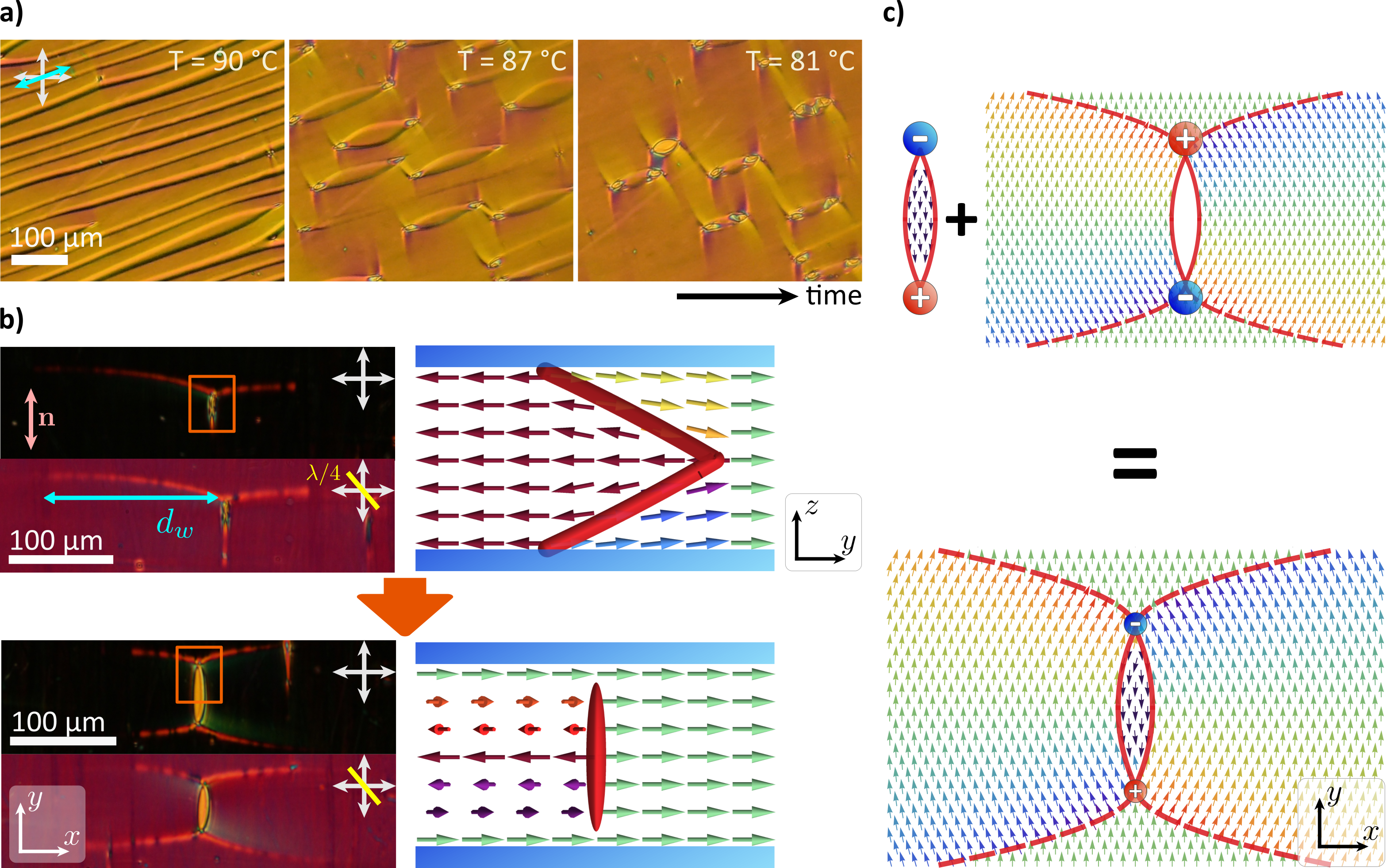}
\caption{\label{fig:fig2} (a) Polarizing optical microscopy (POM) images of the evolution of ferroelectric domains near the $\NS-\NF$ transition while cooling with a cooling rate of 2 K/min. (b) POM images of ferroelectric domains in a 5 $\mu$m thick LC cell with parallel rubbing in different optical configurations with $d_w$ denoting the extent of the wing deformation. A transformation from a uniform (top) to a twisted domain (bottom) is visible. A cross-sectional schematic of the polarization field in the region highlighted by the orange rectangles, before and after the structural transformation, is shown on the right. (c) Schematics illustrating how the bound charges generated by the embedded oppositely polarized domain (upper left) are screened by the wing-like parabolic deformation of the polarization field in the surrounding domain (upper right), resulting in an overall reduction of the bound charge.}
\end{figure*}

\begin{figure*}
\includegraphics[width=1\textwidth]{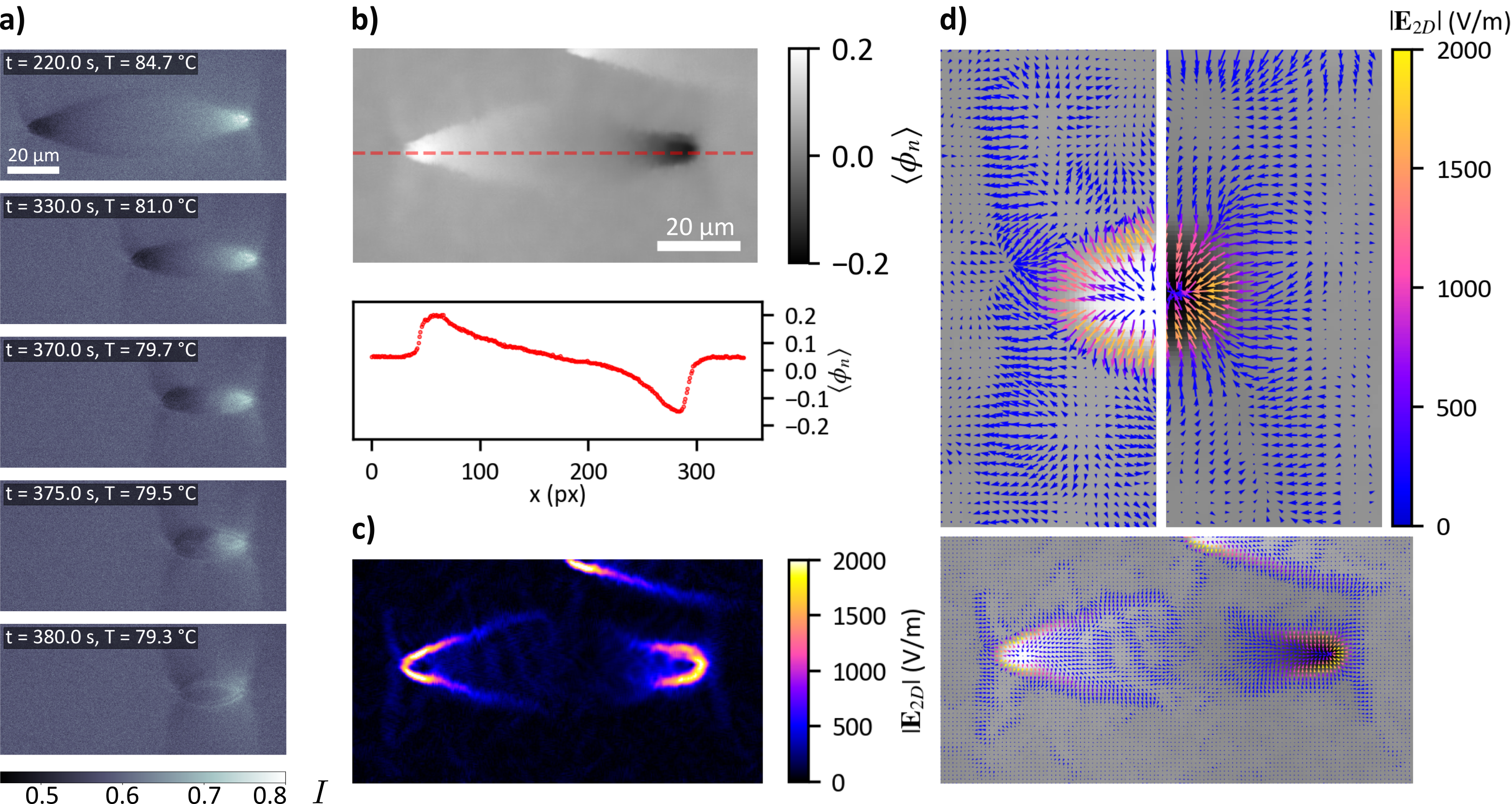}
\caption{\label{fig:fig3} (a) Fluorescence images of the time evolution of the domain shrinking in 5 $\mu$m parallel rubbed polyimide cells with F7 + 0.008 wt\% mRhB mixture upon cooling from $92\, ^{\circ}$C to $79\, ^{\circ}$C at 2 K/min. (b) Normalized 2D potential $\langle\phi_n\rangle$ (Equation (1)), calculated from the fluorescence microscopy image of a ferroelectric domain in a 5 $\mu$m parallel rubbed cell. A line profile along the dashed line is plotted underneath. (c) The magnitude of the local electric field $|\langle \mathbf{E}_{2D}\rangle|$, numerically evaluated using Equation (2). (d) Plot of a local electric field $\langle \mathbf{E}_{2D}\rangle$ overlaid on the normalized potential $\langle\phi_n\rangle$.}.
\end{figure*}

\begin{figure*}
\includegraphics[width=1\textwidth]{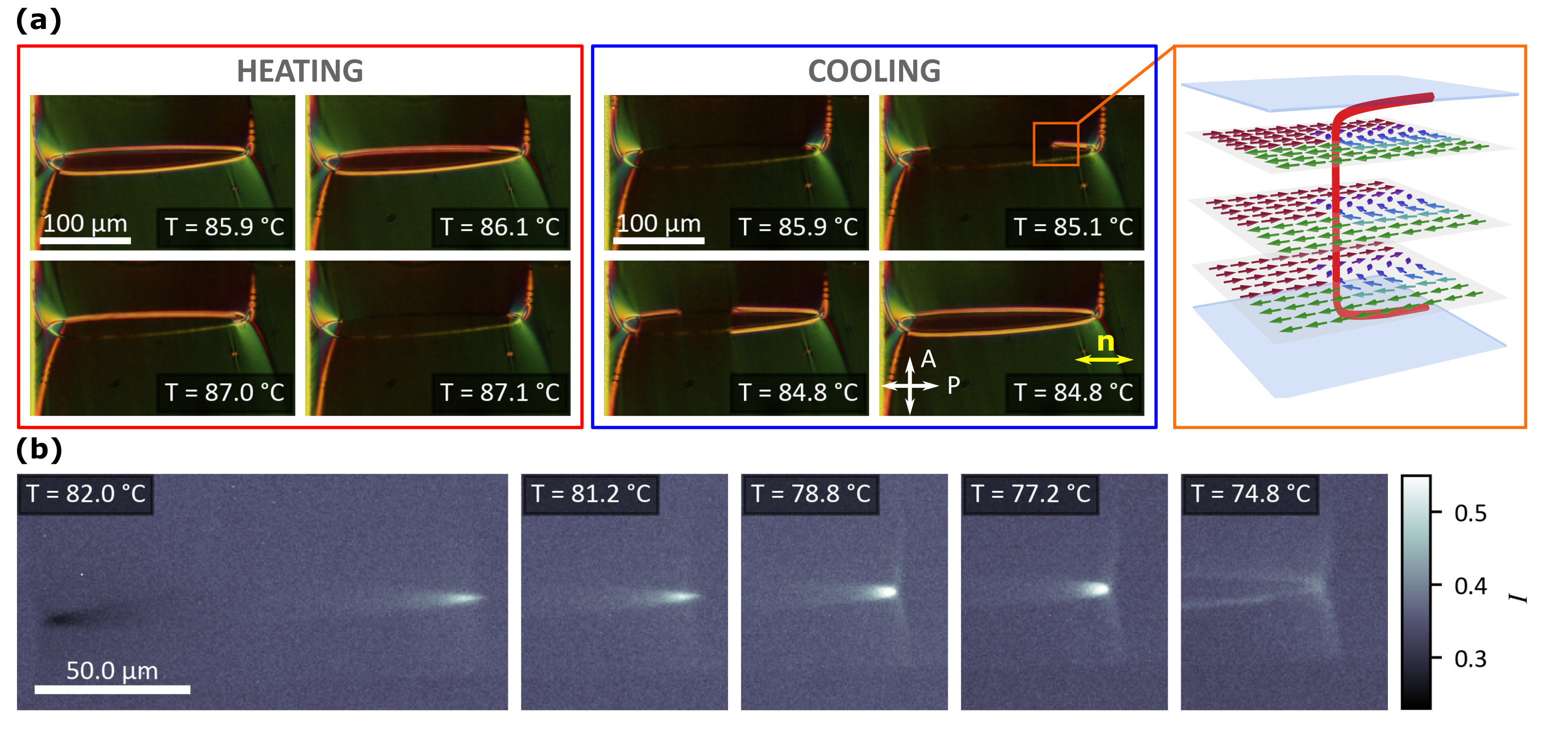}
\caption{\label{fig:fig4} a) Polarizing optical microscopy (POM) images of a FNL in a custom-made 5 $\mu$m thick photopatterned LC cell, showing domain wall evolution during cooling and heating in the ferroelectric nematic phase. A schematic of the boundary between the Ising and Neel-type wall is depicted on the right. The red line denotes a half-integer disclination line, light blue squares represent confining glass surfaces, and the arrows show the orientation of the polarization. b) Fluorescence images depicting the ferroelectric domain evolution during cooling in a 5 $\mu$m photoaligned LC cell with an F7 + 0.008 wt\% mRhB mixture. 
}
\end{figure*}

\textit{Results}---Typically, the $\NF$ phase is preceded by the antiferroelectric splay nematic phase ($\NS$, also refered to as M \cite{nishikawa_fluid_2017}, $\mathrm{N}_x$, $\mathrm{N}_{AF}$ and $\mathrm{SmZ_A}$ \cite{chen_smectic_2023} phase), which has a periodic structure of splayed domains with opposite polarizations separated by Ising walls \cite{mertelj_splay_2018, rosseto_theory_2020, medle_rupnik_antiferroelectric_2025}. At the transition to the $\NF$ phase, homogeneous elongated domains with opposite polarization start to grow. At the sample-air boundary, a triangular domain connects the opposite domains. In Fig.~\ref{fig:fig1}, polarizing optical microscopy (POM) and FM images of the domain structure in a LC cell without surface treatment filled with the mixture F7 doped with 0.008 wt\% of mRhB are shown for two cases; the first, at 70 °C shows the domain structure that forms during cooling from the antiferroelectric phase, the second one is the equilibrated structure at 40 °C. While the POM images look similar in both cases, the FM images show a clear difference. At 40 °C, the domains look homogeneous, the domain walls appear as slightly asymmetric double bright lines, and the sample-air boundary clearly shows an inhomogeneous distribution of cations. At 70 °C, the concentration of ions is less homogeneous, the domain walls appear as a pair of dark and bright lines, and the cations at the sample-air boundary are more evenly distributed than at 40 °C. 

Fig.~\ref{fig:fig2}(a) illustrates the evolution of the domains in LC cells with parallel rubbing upon cooling. Unlike the previous cell without surface treatment, where both signs of $\mathbf{P}$ were equally likely, here the surface favors one sign of $\mathbf{P}$. Initially, at the transition from the $\NS$ phase, elongated domains with both signs of $\mathbf{P}$ develop. Later, the domains disfavored by the surface become much thinner, but they can still be very long (several hundred micrometers). At the point-shaped tips, wing-like director deformation with parabolic shape extends into the surrounding opposite domain (Fig.~\ref{fig:fig2}(b)). Apart from the tips, the domain walls are optically invisible, indicating that they remain of Ising-type. Over time, the unfavoured domains annihilate. This process begins at each domain tip, where a twisted domain nucleates, then it propagates through the domain. In POM between crossed polarizers, these in-plane twist domains appear as bright yellow regions (Fig.~\ref{fig:fig2}(b)). This process is accompanied by a decrease in domain length and subsequent annihilation (see Movie 1 in Supplemental material \cite{supplemental}). Fig.~\ref{fig:fig3}(a) shows FM images of the annihilation process. The dark and bright areas around the domains' tips show that the tips act as sources of positive and negative electrostatic potential. During cooling, the potential difference increases, making the structure progressively more unstable. Eventually, it reaches a critical instability, at which point the twist domain suddenly expands. In FM images, this process results in a more homogeneous ion distribution (Fig.~\ref{fig:fig3}(a)), indicating that the creation of such more complex domain structure reduces the electrostatic potential. In Fig.~\ref{fig:fig3}(b) the potential is evaluated numerically using Equation (1). Fig.~\ref{fig:fig3}(c-d) show the average 2D local field around a domain tip as calculated by
\begin{equation} 
\begin{split}
\langle \mathbf{E}_{2\mathrm{D}} \rangle
=
-\frac{k_{\mathrm{B}}T}{e}\,
\nabla_{xy}\langle \phi_n \rangle
\end{split}
\label{eq2}
\end{equation}
(see End Matter for details). It should be noted that the actual three-dimensional local electric field, $\mathbf{E}(x,y,z)$, may be considerably larger than $\langle \mathbf{E}_{2\mathrm{D}} \rangle$. Nevertheless, $\langle \mathbf{E}_{2\mathrm{D}} \rangle$  provides a useful estimate of the magnitude and spatial distribution of the electrostatic forces and torques within the material.

Fig.~\ref{fig:fig4} shows the evolution of domain walls in the LC cell with photoalignment, which provides nonpolar anchoring. Also in this case, opposite domains end with pointed tips, but they do not annihilate. Upon cooling, a transformation of the domain walls from the Ising type to a more complex structure is observed (Fig.~\ref{fig:fig4}(a)). This process is reversible; when the sample is heated, the topological wall returns to the Ising form at approximately the same temperatures at which the wall formed during cooling. This transformation is nucleated at the tips of the domains, where the electrostatic potential increases upon cooling, as can be deduced from the increasing brightness/darkness of the tips in FM images (Fig.~\ref{fig:fig4}(b)). After the transformation, the fluorescence intensity contrast at the tips decreases, indicating a reduction in the electrostatic potential there.

\textit{Discussion}---The tendency to minimize electrostatic self-energy is the reason for domain formation in ferroelectric materials. The bound charges $\rho_b$ are the sources of the so-called depolarization field and this self-energy. They can be expressed as a sum of surface charges $\sigma_s = \boldsymbol{\upsilon}_s \cdot \mathbf{P}$, where $\boldsymbol{\upsilon}_s$  is the surface normal, and volume charges $\rho_V = -\nabla \cdot \mathbf{P}$. Ideally, the electrostatic energy is zero when the \textbf{P} field is everywhere parallel to the surface of the ferroelectric body, while inside the body, the \textbf{P} field adapts to the surface in such a way that only bend and twist deformations are present, and/or domains are separated by chargeless domain walls of zero thickness. In reality, other contributions to the energy play a role, and the final structure is a balance among them all. In ferroelectric nematic liquids, the main additional contributions are orientational and polarization elastic energies and, at the surface, the surface anchoring energy. These contributions cause the domain walls to have finite thickness, and consequently, a depolarization field is located around them. FM images in Fig.~\ref{fig:fig1} show exactly this; the ions are inhomogeneously distributed mostly around the domains’ borders, while inside the domains the distribution is homogeneous, implying there is no local field. The FM images, in particular that at 40 °C, also show a higher ion concentration in the cores of disclination lines and at the sample–air interface. This observation is consistent with ion partitioning effects, which are assumed to increase the concentrations of both positive and negative ions in these regions. While along the disclination lines the ion concentration is homogeneous, at the sample-air interface it reflects the inhomogeneity of the depolarization field caused by the domain walls. Such a distribution is in qualitative agreement with a simple model, in which it is assumed that uniform domains are separated by domain walls of a finite thickness (Fig.~\ref{fig:fig1}). 

Parabolic domain walls have been observed in ferroelectric liquid layers at the air–liquid interface \cite{kumari_ferroelectric_2023}, which mimics a two-dimensional system free from boundary-imposed orientation. These structures were successfully explained by theory, which shows that in 2D, a parabolic domain wall must separate two divergence-free states \cite{kumari_conic_2024}. Here, we look at the cases beyond this 2D approximation. As shown in Fig. ~\ref{fig:fig2}(c), the tips of elongated unfavorable domains act as separated point-like charges. The surrounding polarization screens these charges by adopting a configuration with nonzero $-\nabla \cdot \mathbf{P}$. Fig.~\ref{fig:fig2}(c) shows a schematic of this structure, inferred from POM analysis. The bound charge density of such a structure is mainly concentrated near the tips. As deduced from POM analysis of the wing-like structures, there is an out-of-plane splay deformation around the central part of the parabolic walls, with nonzero $-\nabla \cdot \mathbf{P}$ (End Matter, Fig. \ref{fig:fig5EM}(c)). This indicates that parabolic walls also carry some bound charge. The FM images reveal that, despite this screening, a local field of the order of a few 1000 V/m remains around the tips (Fig.~\ref{fig:fig3}), which ultimately leads to domain annihilation. Interestingly, although annihilation could in principle occur via domain thinning and melting into the apolar nematic phase, this pathway is very rarely observed. Instead, the wing-like deformed structure appears to stabilize the bound charge and suppress this annihilation route. 

Although the Ising domain walls, when parallel to the polarization, are electrostatically neutral, they are not observed deep in the $\NF$ phase. Instead there, the domain walls between uniform domains are of Néel type with two half-integer disclination lines at the layer surfaces. A comparison of the energetics of Ising and Néel domain walls within a simple one-dimensional model shows that increasing \textbf{P} makes Ising walls less energetically favorable than Néel walls, thereby explaining this observation (End Matter). The transformation is of nucleation type, the Néel wall propagates from the nucleation sites, which are typically domain tips. Between an Ising and Néel wall there must be a one half-integer disclination line (Fig.~\ref{fig:fig4}(a)). This half-integer line extends along both confining surfaces to accommodate bend deformation of the Neel wall with uniform alignment at the surfaces.

\textit{Conclusions}---We used fluorescent ions to experimentally demonstrate that local electric fields are localized around domain walls and at the sample surface. Point-like bound charges at the tips of domains are effectively compensated and stabilized by the splayed wing-like structure of the polarization field in the surrounding domain. Upon cooling, the Ising walls that initially separate domains with opposite polarization transform into Néel-type domain walls, with regions of enhanced local electric fields serving as nucleation sites for this transition. Furthermore, the increased ion concentration observed within disclination cores indicates enhanced ion solvation in these regions. Overall, these direct experimental observations of local fields in a ferroelectric liquid establish the role of electrostatic effects in governing domain formation and the structure and topology of ferroelectric domain walls, providing a foundation for understanding and controlling electrostatic interactions in soft ferroelectric materials. 

\textit{Acknowledgments}---A.S., N.O., N.S. and A.M. acknowledge the support of the Slovenian Research and Innovation Agency (Grant Nos. P1-0192, J1-50004, and BI-VB/25-27-011). J.H., C.J.G. and R.J.M. acknowledge funding from UKRI via a Future Leaders Fellowship, grant No. MR/W006391/1.

\textit{Author contributions}---A.M. conceived the hypothesis, designed the study and supervised the project. A.S. performed the experiments. N.O. and N.S. advised on the experimental design and methodology. A.S. and A.M. analyzed the data. A.M. carried out the model calculations. R.J.M., J.H., and C.J.G. provided expert advice on the materials and the research. A.M. and A.S. wrote the first draft of the manuscript. All authors reviewed, revised, and approved the final version of the manuscript.

Data availability: The data that support the findings of this article are openly available at Ref. \cite{sterle_data_2026}.

\bibliography{REFERENCES}

@article{mandle_rational_2017,
	title = {Rational {Design} of {Rod}-{Like} {Liquid} {Crystals} {Exhibiting} {Two} {Nematic} {Phases}},
	volume = {23},
	issn = {1521-3765},
	url = {http://onlinelibrary.wiley.com/doi/10.1002/chem.201702742/abstract},
	doi = {10.1002/chem.201702742},
	number = {58},
	urldate = {2018-02-02},
	journal = {Chemistry – A European Journal},
	author = {Mandle, Richard J. and Cowling, Stephen J. and Goodby, John W.},
	month = oct,
	year = {2017},
	pages = {14554--14562},
}

@article{mertelj_splay_2018,
	title = {Splay {Nematic} {Phase}},
	volume = {8},
	url = {https://link.aps.org/doi/10.1103/PhysRevX.8.041025},
	doi = {10.1103/PhysRevX.8.041025},
	number = {4},
	urldate = {2018-11-13},
	journal = {Physical Review X},
	author = {Mertelj, Alenka and Cmok, Luka and Sebastián, Nerea and Mandle, Richard J. and Parker, Rachel R. and Whitwood, Adrian C. and Goodby, John W. and Čopič, Martin},
	month = nov,
	year = {2018},
	pages = {041025},
}

@article{nishikawa_fluid_2017,
	title = {A {Fluid} {Liquid}-{Crystal} {Material} with {Highly} {Polar} {Order}},
	volume = {29},
	copyright = {© 2017 WILEY‐VCH Verlag GmbH \& Co. KGaA, Weinheim},
	issn = {1521-4095},
	url = {https://onlinelibrary.wiley.com/doi/abs/10.1002/adma.201702354},
	doi = {10.1002/adma.201702354},
	number = {43},
	urldate = {2018-12-18},
	journal = {Advanced Materials},
	author = {Nishikawa, Hiroya and Shiroshita, Kazuya and Higuchi, Hiroki and Okumura, Yasushi and Haseba, Yasuhiro and Yamamoto, Shin-ichi and Sago, Koki and Kikuchi, Hirotsugu},
	year = {2017},
	pages = {1702354},
}

@article{rosseto_theory_2020,
	title = {Theory of the splay nematic phase: {Single} versus double splay},
	volume = {101},
	shorttitle = {Theory of the splay nematic phase},
	url = {https://link.aps.org/doi/10.1103/PhysRevE.101.052707},
	doi = {10.1103/PhysRevE.101.052707},
	number = {5},
	urldate = {2020-07-10},
	journal = {Physical Review E},
	publisher = {American Physical Society},
	author = {Rosseto, Michely P. and Selinger, Jonathan V.},
	month = may,
	year = {2020},
	pages = {052707},
}

@article{chen_polar_2021,
	chapter = {Physical Sciences},
	title = {Polar in-plane surface orientation of a ferroelectric nematic liquid crystal: {Polar} monodomains and twisted state electro-optics},
	volume = {118},
	copyright = {Copyright © 2021 the Author(s). Published by PNAS.. https://creativecommons.org/licenses/by-nc-nd/4.0/This open access article is distributed under Creative Commons Attribution-NonCommercial-NoDerivatives License 4.0 (CC BY-NC-ND).},
	issn = {0027-8424, 1091-6490},
	shorttitle = {Polar in-plane surface orientation of a ferroelectric nematic liquid crystal},
	url = {https://www.pnas.org/content/118/22/e2104092118},
	doi = {10.1073/pnas.2104092118},
	number = {22},
	urldate = {2021-05-31},
	journal = {Proceedings of the National Academy of Sciences},
	publisher = {National Academy of Sciences},
	author = {Chen, Xi and Korblova, Eva and Glaser, Matthew A. and Maclennan, Joseph E. and Walba, David M. and Clark, Noel A.},
	month = jun,
	year = {2021},
}

@article{caimi_surface_2021,
	title = {Surface alignment of ferroelectric nematic liquid crystals},
	volume = {17},
	issn = {1744-6848},
	url = {https://pubs.rsc.org/en/content/articlelanding/2021/sm/d1sm00734c},
	doi = {10.1039/D1SM00734C},
	urldate = {2021-08-23},
	journal = {Soft Matter},
	publisher = {The Royal Society of Chemistry},
	author = {Caimi, Federico and Nava, Giovanni and Barboza, Raouf and Clark, Noel A. and Korblova, Eva and Walba, David M. and Bellini, Tommaso and Lucchetti, Liana},
	month = aug,
	year = {2021},
	pages = {8130--8139},
}

@article{everts_ionically_2021,
	title = {Ionically {Charged} {Topological} {Defects} in {Nematic} {Fluids}},
	volume = {11},
	url = {https://link.aps.org/doi/10.1103/PhysRevX.11.011054},
	doi = {10.1103/PhysRevX.11.011054},
	number = {1},
	urldate = {2022-01-03},
	journal = {Physical Review X},
	publisher = {American Physical Society},
	author = {Everts, Jeffrey C. and Ravnik, Miha},
	month = mar,
	year = {2021},
	pages = {011054},
}

@article{kumari_ferroelectric_2023,
	title = {Ferroelectric nematic liquids with conics},
	volume = {14},
	copyright = {2023 The Author(s)},
	issn = {2041-1723},
	url = {https://www.nature.com/articles/s41467-023-36326-1},
	doi = {10.1038/s41467-023-36326-1},
	number = {1},
	urldate = {2023-02-15},
	journal = {Nature Communications},
	publisher = {Nature Publishing Group},
	author = {Kumari, Priyanka and Basnet, Bijaya and Wang, Hao and Lavrentovich, Oleg D.},
	month = feb,
	year = {2023},
	note = {Number: 1},
	pages = {748},
}

@article{sebastian_polarization_2023,
	title = {Polarization patterning in ferroelectric nematic liquids via flexoelectric coupling},
	volume = {14},
	copyright = {2023 The Author(s)},
	issn = {2041-1723},
	url = {https://www.nature.com/articles/s41467-023-38749-2},
	doi = {10.1038/s41467-023-38749-2},
	number = {1},
	urldate = {2023-05-26},
	journal = {Nature Communications},
	publisher = {Nature Publishing Group},
	author = {Sebastián, Nerea and Lovšin, Matija and Berteloot, Brecht and Osterman, Natan and Petelin, Andrej and Mandle, Richard J. and Aya, Satoshi and Huang, Mingjun and Drevenšek-Olenik, Irena and Neyts, Kristiaan and Mertelj, Alenka},
	month = may,
	year = {2023},
	note = {Number: 1},
	pages = {3029},
}

@article{caimi_fluid_2023,
	title = {Fluid superscreening and polarization following in confined ferroelectric nematics},
	copyright = {2023 The Author(s), under exclusive licence to Springer Nature Limited},
	issn = {1745-2481},
	url = {https://www.nature.com/articles/s41567-023-02150-z},
	doi = {10.1038/s41567-023-02150-z},
	urldate = {2023-08-16},
	journal = {Nature Physics},
	publisher = {Nature Publishing Group},
	author = {Caimi, Federico and Nava, Giovanni and Fuschetto, Susanna and Lucchetti, Liana and Paiè, Petra and Osellame, Roberto and Chen, Xi and Clark, Noel A. and Glaser, Matthew A. and Bellini, Tommaso},
	month = jul,
	year = {2023},
	pages = {1--9},
}

@article{mathe_electrically_2024,
	title = {Electrically activated ferroelectric nematic microrobots},
	volume = {15},
	copyright = {2024 The Author(s)},
	issn = {2041-1723},
	url = {https://www.nature.com/articles/s41467-024-50226-y},
	doi = {10.1038/s41467-024-50226-y},
	number = {1},
	urldate = {2024-08-21},
	journal = {Nature Communications},
	publisher = {Nature Publishing Group},
	author = {Máthé, Marcell Tibor and Nishikawa, Hiroya and Araoka, Fumito and Jákli, Antal and Salamon, Péter},
	month = aug,
	year = {2024},
	note = {Number: 1},
	pages = {6928},
}

@article{mathe_electric_2023,
	title = {Electric field-induced interfacial instability in a ferroelectric nematic liquid crystal},
	volume = {13},
	copyright = {2023 The Author(s)},
	issn = {2045-2322},
	url = {https://www.nature.com/articles/s41598-023-34067-1},
	doi = {10.1038/s41598-023-34067-1},
	number = {1},
	urldate = {2024-09-17},
	journal = {Scientific Reports},
	publisher = {Nature Publishing Group},
    author = {M{\'a}th{\'e}, Marcell Tibor and Farkas, Bendeg{\'u}z and P{\'e}ter, L{\'a}szl{\'o} and Buka, {\'A}gnes and J{\'a}kli, Antal and Salamon, P{\'e}ter},
	month = apr,
	year = {2023},
	note = {Number: 1},
	pages = {6981},
}

@article{barboza_explosive_2022,
	title = {Explosive electrostatic instability of ferroelectric liquid droplets on ferroelectric solid surfaces},
	volume = {119},
	url = {https://www.pnas.org/doi/10.1073/pnas.2207858119},
	doi = {10.1073/pnas.2207858119},
	number = {32},
	urldate = {2022-08-13},
	journal = {Proceedings of the National Academy of Sciences},
	publisher = {Proceedings of the National Academy of Sciences},
	author = {Barboza, Raouf and Marni, Stefano and Ciciulla, Fabrizio and Mir, Farooq Ali and Nava, Giovanni and Caimi, Federico and Zaltron, Annamaria and Clark, Noel A. and Bellini, Tommaso and Lucchetti, Liana},
	month = aug,
	year = {2022},
	pages = {e2207858119},
}

@article{yi_chiral_2024,
	title = {Chiral $\pi$ domain walls composed of twin half-integer surface disclinations in ferroelectric nematic liquid crystals},
	volume = {121},
	url = {https://www.pnas.org/doi/10.1073/pnas.2413879121},
	doi = {10.1073/pnas.2413879121},
	number = {52},
	urldate = {2025-02-04},
	journal = {Proceedings of the National Academy of Sciences},
	publisher = {Proceedings of the National Academy of Sciences},
	author = {Yi, Shengzhu and Hong, Zening and Ma, Zhongjie and Zhou, Chao and Jiang, Miao and Huang, Xiang and Huang, Mingjun and Aya, Satoshi and Zhang, Rui and Wei, Qi-Huo},
	month = dec,
	year = {2024},
	pages = {e2413879121},
}

@article{medle_rupnik_antiferroelectric_2025,
	title = {Antiferroelectric {Order} in {Nematic} {Liquids}: {Flexoelectricity} {Versus} {Electrostatics}},
	volume = {12},
	copyright = {© 2025 The Author(s). Advanced Science published by Wiley-VCH GmbH},
	issn = {2198-3844},
	shorttitle = {Antiferroelectric {Order} in {Nematic} {Liquids}},
	url = {https://onlinelibrary.wiley.com/doi/abs/10.1002/advs.202414818},
	doi = {10.1002/advs.202414818},
	number = {9},
	urldate = {2025-03-11},
	journal = {Advanced Science},
	author = {Medle Rupnik, Peter and Hanžel, Ema and Lovšin, Matija and Osterman, Natan and Gibb, Calum Jordan and Mandle, Richard J. and Sebastián, Nerea and Mertelj, Alenka},
	year = {2025},
	pages = {2414818},
}

@article{sterle_light-induced_2023,
	title = {Light-induced dynamics of liquid-crystalline droplets on the surface of iron-doped lithium niobate crystals},
	volume = {13},
	copyright = {© 2022 Optica Publishing Group},
	issn = {2159-3930},
	url = {https://opg.optica.org/ome/abstract.cfm?uri=ome-13-1-282},
	doi = {10.1364/OME.477717},
	number = {1},
	urldate = {2025-03-18},
	journal = {Optical Materials Express},
	publisher = {Optica Publishing Group},
	author = {Sterle, Anej and Cmok, Luka and Sebastián, Nerea and Mertelj, Alenka and Kong, Yongfa and Zhang, Xinzheng and Drevenšek-Olenik, Irena},
	month = jan,
	year = {2023},
	pages = {282--294},
}

@article{gibb_spontaneous_2024,
	title = {Spontaneous symmetry breaking in polar fluids},
	volume = {15},
	copyright = {2024 The Author(s)},
	issn = {2041-1723},
	url = {https://www.nature.com/articles/s41467-024-50230-2},
	doi = {10.1038/s41467-024-50230-2},
	number = {1},
	urldate = {2025-03-25},
	journal = {Nature Communications},
	publisher = {Nature Publishing Group},
	author = {Gibb, Calum J. and Hobbs, Jordan and Nikolova, Diana I. and Raistrick, Thomas and Berrow, Stuart R. and Mertelj, Alenka and Osterman, Natan and Sebastián, Nerea and Gleeson, Helen F. and Mandle, Richard J.},
	month = jul,
	year = {2024},
	pages = {5845},
}

@article{kumari_conic_2024,
	title = {Conic sections in ferroelectric nematics: {Experiments} and mathematical modeling},
	volume = {6},
	shorttitle = {Conic sections in ferroelectric nematics},
	url = {https://link.aps.org/doi/10.1103/PhysRevResearch.6.043207},
	doi = {10.1103/PhysRevResearch.6.043207},
	number = {4},
	urldate = {2025-11-17},
	journal = {Physical Review Research},
	publisher = {American Physical Society},
	author = {Kumari, Priyanka and Kurochkin, Olexandr and Nazarenko, Vassili G. and Lavrentovich, Oleg D. and Golovaty, Dmitry and Sternberg, Peter},
	month = nov,
	year = {2024},
	pages = {043207},
}

@article{lavrentovich_twist_2025,
	title = {Twist, splay, and uniform domains in ferroelectric nematic liquid crystals},
	volume = {16},
	copyright = {2025 The Author(s)},
	issn = {2041-1723},
	url = {https://www.nature.com/articles/s41467-025-61840-9},
	doi = {10.1038/s41467-025-61840-9},
	number = {1},
	urldate = {2025-11-17},
	journal = {Nature Communications},
	publisher = {Nature Publishing Group},
	author = {Lavrentovich, Maxim O. and Kumari, Priyanka and Lavrentovich, Oleg D.},
	month = jul,
	year = {2025},
	pages = {6516},
}

@article{cmok_running_2023,
	title = {Running streams of a ferroelectric nematic liquid crystal on a lithium niobate surface},
	volume = {50},
	issn = {0267-8292},
	url = {https://doi.org/10.1080/02678292.2022.2161018},
	doi = {10.1080/02678292.2022.2161018},
	number = {7-10},
	urldate = {2026-02-18},
	journal = {Liquid Crystals},
	publisher = {Taylor \& Francis},
	author = {Cmok, Luka and Coda, Virginie and Sebastian, Nerea and Mertelj, Alenka and Zgonik, Marko and Aya, Satoshi and Huang, Mingjun and Montemezzani, Germano and Drevensek-Olenik, Irena},
	month = aug,
	year = {2023},
	note = {\_eprint: https://doi.org/10.1080/02678292.2022.2161018},
	pages = {1478--1485},
}

@misc{petelin_dtmm_2020,
	title = {{IJSComplexMatter}/dtmm: {Version} 0.6.1},
	copyright = {Open Access},
	shorttitle = {{IJSComplexMatter}/dtmm},
	url = {https://zenodo.org/record/1245999},
	doi = {10.5281/ZENODO.1245999},
	urldate = {2026-07-16},
	publisher = {Zenodo},
	author = {Petelin, Andrej},
	month = nov,
	year = {2020},
}

@misc{erkoreka_flexoelectricity-driven_2026,
	title = {Flexoelectricity-driven softening of bend elasticity leads to spontaneous chiral symmetry breaking in a polar fluid},
	url = {http://arxiv.org/abs/2602.15687},
	doi = {10.48550/arXiv.2602.15687},
	urldate = {2026-06-09},
	publisher = {arXiv},
	author = {Erkoreka, Aitor and Martinez-Perdiguero, Josu and Cmok, Luka and Hanžel, Ema and Hobbs, Jordan and Gibb, Calum J. and Mandle, Richard J. and Sebastián, Nerea and Mertelj, Alenka},
	month = feb,
	year = {2026},
	note = {arXiv:2602.15687 [cond-mat.soft]},	
}

@article{chen_smectic_2023,
	title = {The smectic {ZA} phase: {Antiferroelectric} smectic order as a prelude to the ferroelectric nematic},
	volume = {120},
	shorttitle = {The smectic {ZA} phase},
	url = {https://www.pnas.org/doi/10.1073/pnas.2217150120},
	doi = {10.1073/pnas.2217150120},
	number = {8},
	urldate = {2023-03-20},
	journal = {Proceedings of the National Academy of Sciences},
	publisher = {Proceedings of the National Academy of Sciences},
	author = {Chen, Xi and Martinez, Vikina and Korblova, Eva and Freychet, Guillaume and Zhernenkov, Mikhail and Glaser, Matthew A. and Wang, Cheng and Zhu, Chenhui and Radzihovsky, Leo and Maclennan, Joseph E. and Walba, David M. and Clark, Noel A.},
	month = feb,
	year = {2023},
	pages = {e2217150120},	
}

@misc{sterle_data_2026,
author = {{A. Sterle \textit{et al.}}},
year = {2026},
doi = {10.5281/zenodo.21394238},
note= {10.5281/zenodo.21394238}
}

@misc{supplemental,
  note = {See Supplemental Material at [URL] for a video of domain annihilation in polymide cells.},
  year = {2026},
}

\onecolumngrid
\section*{End Matter}
\twocolumngrid

\textit{Materials}---In experiments, F7 mixture (30:70 molar ratio of Compound 1 from Ref  \cite{gibb_spontaneous_2024} and material DIO \cite{nishikawa_fluid_2017}) with the phase sequence N 100°C $\NS$ 91°C $\NF$ was used. Rhodamine B was modified by salt metathesis to exchange the chloride counterion (Cl$^-$) for hexafluorophosphate (PF$_6^-$), giving modified Rhodamine B (mRhB). mRhB was dissolved in acetone to prepare a solution with a concentration of 0.51 mg/ ml. An appropriate amount of solution was added to F7 at room temperature to achieve concentrations ranging from 0.004 wt\% to 0.06 wt\% (corresponding to $5\cdot 10^{22}$ to $8\cdot 10^{23}$ ions/$\mathrm{m}^3$). The mixture was heated to the N phase ($\approx 100$ °C) to fully evaporate the acetone. Commercial (EHC) liquid crystal cells with ITO electrodes on both substrates, with or without rubbed polyimide layers, were filled with capillary action in the nematic phase. The thickness of the LC cells was in most cases 5 $\mu\mathrm{m}$.

\textit{Photoalignment}---For photoalignment, custom LC cells were assembled using ITO-coated glass substrates. Substrates were first cleaned and treated with plasma for 15 min to improve surface wettability and coating uniformity. A 1 wt\% solution of Brilliant Yellow in N,N-Dimethylformamide (DMF) was spin-coated onto the substrates at 500 rpm for 30 s, followed by 6000 rpm for 60 s. After spin coating, the films were dried at 120 °C for 30 min. Photoalignment was performed by illuminating the assembled LC cells with polarized light from an LED source (Thorlabs M405LP2) with a peak emission wavelength of 450 nm. The samples were illuminated for 3 min at an intensity of I = 75 mW/$\mathrm{cm}^2$ to achieve planar alignment. 

\textit{Fluorescence microscopy (FM) and calculation of local fields}---FM was performed with a Nikon Eclipse Ti2 microscope with the CoolLED pE-300 multi-LED source. A Chroma mCherry filter cube (Chroma 59022) with the excitation wavelength of $554-589.5$ nm and emission wavelength of $602-666$ nm was used. Images were taken at 10$\times$ (Nikon Plan Fluor) and 20$\times$ (Nikon T Plan SLWD) objectives. The sample was heated and cooled using an Instec HCS402 heating stage controlled by an Instec mK2000 temperature controller. To prevent bleaching, typically, samples were illuminated at lower intensities and recorded at exposure times $\sim1$ s. The fluorescence images of domains were first flat-field corrected by taking the grayscale image and then normalizing it by an image of a a region with a homogenous ion distribution. A non-local means filter from the python scikit-image library (skimage.restoration.denoise\_nl\_means) was then applied for noise reduction and better evaluation of the numerical gradients. Parameters for denoising were the following: h = 0.03, patch\_size = 4, path\_distance = 7. An average local field $\langle \mathbf{E}_{2\mathrm{D}} \rangle$ was then calculated by using the Equation \eqref{eq1} and \eqref{eq2}. 

\begin{figure*}
\includegraphics[width=1\textwidth]{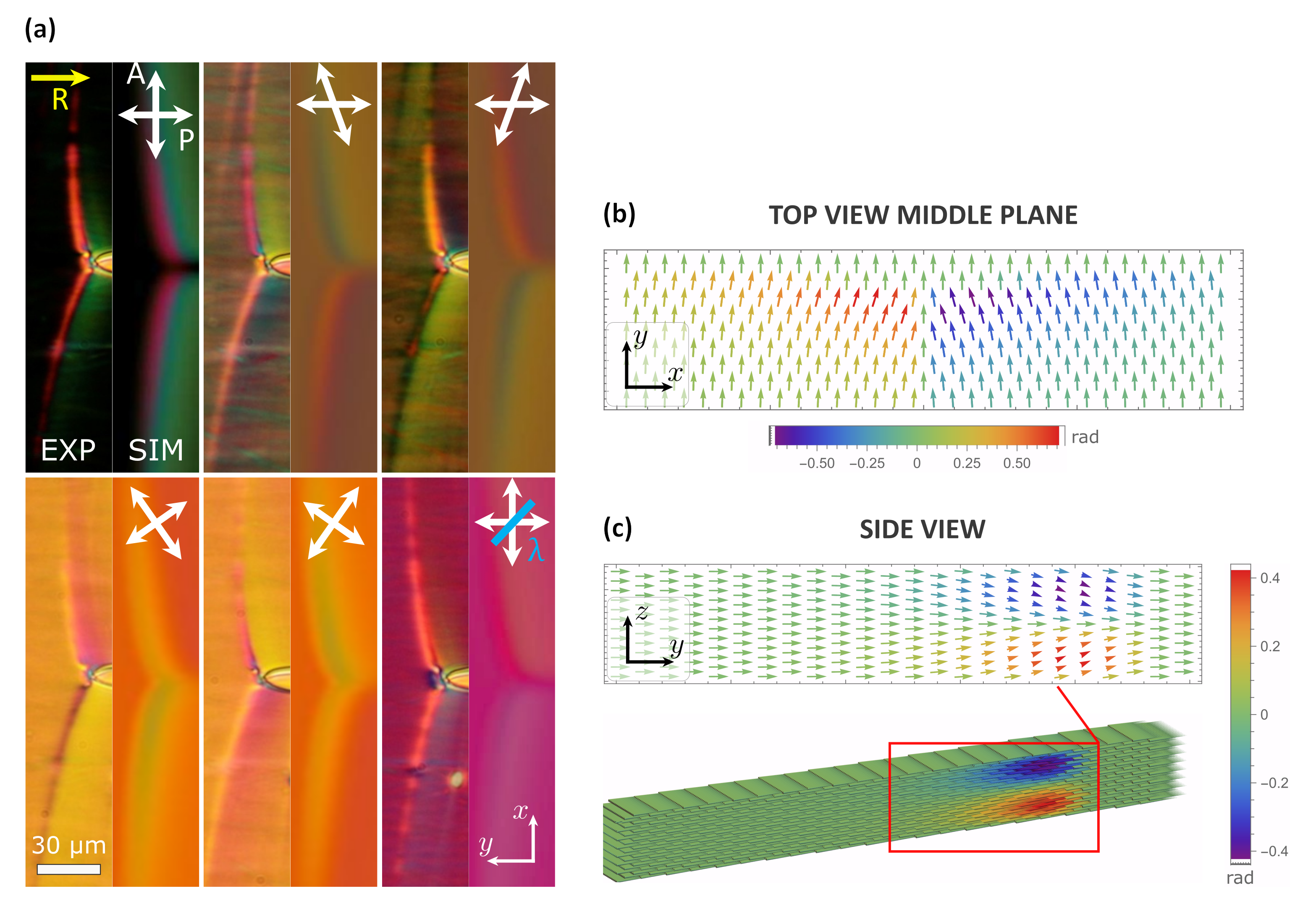}
\caption{\label{fig:fig5EM} (a) Side-by-side comparison of the experimental POM images (EXP) of pure F7 in a 5 $\mu \mathrm{m}$ parallel rubbed cell, and images obtained from optical simulations (SIM) performed using the diffractive transfer matrix method \cite{petelin_dtmm_2020}. The transmission axis of the analyser (A) and the polarizer (P) is denoted with a double-sided arrow. The yellow arrow, labeled R, indicates the rubbing direction of the polyimide. Bottom right image was taken with a full-wave $\lambda$ plate inserted. (b) View of the simulated in-plane polarization structure at the domain tip at $z= 0.5\, d$, $d$ being the cell thickness. (c) Side view ($yz$ plane) of the simulated polarization structure, showcasing the out-of-plane splay deformation, responsible for the color shift along the parabolic walls.}
\end{figure*}

\textit{Polarizing microscopy (POM) analysis}---
Polarizing microscopy experiments were performed using a Nikon Optiphot-2 POL microscope. By systematically varying the orientations of the polarizer, analyzer, and sample, as well as introducing additional optical elements, including a $\lambda$ plate, images were acquired for a range of optical configurations. Images and videos were captured using a Canon EOS M200 camera. The principal characteristics of the structures were inferred from the POM images. The resulting director field was then used as input to the diffractive transfer matrix method (dtmm) \cite{petelin_dtmm_2020} to simulate images corresponding to the experimental optical configurations. The green shadows in POM images with crossed polarizers, which accompany the wings, show that the \textbf{n} field goes from the uniform alignment at the glass surfaces to the splayed structure (Fig.~\ref{fig:fig5EM}(b)) in the middle of the cell. The red color along the central part of the parabolic walls indicates an out-of-plane splay deformation as depicted in Fig.~\ref{fig:fig5EM}(c).

\textit{Comparison of free energy cost between Ising and Néel domain walls}---To estimate energy cost of the Ising and Néel domain walls a Landau - de Gennes type of free energy is assumed:
\begin{eqnarray}
F = \int \Bigg[&
\frac{1}{2}K_1(\nabla\cdot\mathbf{n})^2
+
\frac{1}{2}K_3
\left|\mathbf{n}\times(\nabla\times\mathbf{n})\right|^2
\nonumber \\ \nonumber
&+
\frac{1}{4}B(P^2-P_0^2)^2
+
\frac{1}{2}K_P(\nabla P)^2
\\ 
&-
\frac{1}{2}(\nabla\cdot\mathbf{P})\phi
\Bigg]\,dV.
\end{eqnarray}
Here, $P_0$ is the equilibrium magnitude of spontaneous polarization in bulk; the polarization is assumed to be parallel to the director $\mathbf{n}$, $\mathbf{P}=P\,\mathbf{n}$; $K_1$ and $K_3$ are the splay and bend elastic constants, respectively, $K_P$ is the elastic constant associated with gradients of the magnitude of polarization, $B$ is the Landau coefficient, and $\phi$ electrostatic potential given by the Poisson equation, $\varepsilon \varepsilon_0\nabla^2\phi=-\nabla\cdot\mathbf{P}$. If the structure of the Ising wall is assumed to be $\mathbf{P} =P(x)(0,1,0)$, where $P$ linearly changes from $-P_0$ to $P_0$ within the domain wall of the thickness $d_{\mathrm{dwI}}$, elsewhere it is constant and equal to $P_0$, the domain wall energy of a wall of an area $A$ is $F_{Ising}=(4A\sqrt{BK_P}P_0^3)/\sqrt{15}$ and $d_{\mathrm{dwI}}=\sqrt{15K_P}/(\sqrt{B}P_0)$. Similarly, if the structure of the N\'eel wall is assumed to be $\mathbf{P}=P_0\,(\sin\theta(x),\cos\theta(x),0)$, where $\theta(x)$ linearly changes from 0 to $\pi$ within the domain wall of thickness $d_{\mathrm{dwN}}$, the domain wall energy of a wall with area $A$ is $F_{\mathrm{N\acute{e}el}}=(\pi A\sqrt{K_1+K_3}P_0)/(2\sqrt{\varepsilon \varepsilon_0})$ and its thickness $d_{\mathrm{dwN}}=(\pi\sqrt{(K_1+K_3)\varepsilon \varepsilon_0})/{P_0}$. Comparison of the domain wall energies of these simple structures shows that for $P_0^2>\pi\sqrt{15(K_1+K_3)}/(8\sqrt{BK_P \varepsilon \varepsilon_0})$, N\'eel domain walls are energetically more stable. Assuming $P_0\approx 0.01$As/m$^2$, $K_{1,3} \approx 10^{-11}$ N, $B\approx 1/(\varepsilon\varepsilon_0 P_0^2)$ \cite{medle_rupnik_antiferroelectric_2025}, and $\varepsilon=100$, we can estimate $K_P\approx 10^{-5}$ Nm$^4$/(As)$^2$.

\end{document}